\documentclass[a4paper]{jpconf}
\usepackage{graphicx}
\usepackage{color}
\begin{document}

\title{Equation of State in Non-Zero Magnetic Field}

\author{Nada Ezzelarab, Abdel Magied Diab and Abdel Nasser Tawfik}

\address{Egyptian Center for Theoretical Physics (ECTP), Modern University for Technology and Information (MTI), 11571 Cairo, Egypt}
\address{World Laboratory for Cosmology And Particle Physics (WLCAPP), Cairo, Egypt}

\ead{nada.ezzelarab@eng.mti.edu.eg}

\begin{abstract}
The Polyakov linear-sigma model (PLSM) and Hadron Resonance Gas (HRG) model are considered to study the hadronic and partonic equation(s) of state, the pressure, and response to finite magnetic field, the magnetization. The results are confronted to recent lattice QCD calculations. Both models are in fairly good agreement with the lattice. 
\end{abstract}


In heavy-ion collisions, especially the peripheral ones, the magnetic field likely has significant effects, which can be coupled to experimental observables, for instance, in STAR experiment at the relativistic heavy-ion collider (RHIC), and ALICE experiment at the large hadron collider (LHC). Influences of such huge magnetic fields, ${\cal O}(m_{\pi}^2)$, on the hadronic and partonic matter and on the structure of the quark-gluon phase was analyzed in various models, such as hadron resonance gas (HRG) \cite{HRG1}, and also estimated in lattice QCD simulations \cite{QCD:2013d,lattice:2014}. Coupling Polyakov loops to the linear-sigma model (PLSM) introduces color charge interactions to the pure gauge field. PLSM has revealed interesting features from studying the responses of QCD equation(s) of state to non-zero magnetic field \cite{TN:magnet}. In the present work, we confront our calculations on pressure and magnetization from HRG and PLSM at vanishing and finite magnetic field with recent lattice QCD calculations \cite{lattice:2014}.

The influence of finite magmatic to HRG and PLSM can be divided into two parts.
\begin{itemize} 
\item The first one, the dispersion relation becomes subject of modification \cite{book1,book2}. Assuming that the magnetic field is directed along $z$-axis, then in HRG, 
\begin{equation}
E_{\ell s_{z}} = \left[m^{2}+P_{||}^{2}+2 g_{s} (\ell+s_{z}+1/2) B\right]^{1/2},  \label{disp1}
\end{equation}
and in PLSM,  
\begin{eqnarray}
E_{B, f} (B) =\left[p_{z}^{2}+m_{f}^{2}+|q_{f}|(2n+1-\sigma) B\right]^{1/2}, \label{eqLL}
\end{eqnarray}
where $n$ is Landau quantum number and $\sigma=\pm S/2$ is related to the spin quantum number. 

\item The second part is the phase space which gets modifications due the effects of the magnetic catalysis,
\begin{eqnarray} 
\int \frac{d^3 p}{(2 \pi)^3}~ \longrightarrow~  \frac{|q_{f}| B}{2 \pi} \sum_{\nu=0}^{\infty} \int \frac{d p}{2 \pi} (2-1 \delta_{0\nu}), \label{DR}
\end{eqnarray} 
where $2-1 \delta_{0\nu}$ stands for degenerate Landau levels.
\end{itemize}

Treating hadron resonances as a collision-free gas \cite{Karsch:2003vd,Redlich:2004gp,Tawfik:2004sw,Taw3} allows the estimation of various thermodynamic quantities from the grand canonical partition function \cite{Karsch:2003vd,Redlich:2004gp}. In non-zero magnetic field,
\begin{eqnarray}
\ln Z(T,V,\mu)&=& \frac{V g_i}{(2\pi)^3} \,|Q_i|\,eB_z \sum_n \sum_{s_z} \int_0^{\infty} \pm \,dp_z \ln\left\{1\pm \, \exp[(\mu_i - E_{i}^{B_z})/T]\right\}, \label{eq:PFq1}
\end{eqnarray}
where $\pm$ stands fermions and bosons, respectively, $V$ is the volume and $g_{i}$ is the degeneracy for $i^{th}$ hadron. 

In the mean field approximation of PLSM, the partition function expresses the exchange of energy between particle and antiparticle at temperature ($T$) and baryon chemical potential ($\mu_f$). At finite magnetic field and assuming Landau quantization \cite{LSMmagnetic:2015}, the quark-antiquark potential is given as
\begin{eqnarray}
\Omega_{  \bar{q}q}(T, \mu _f, B) &=& -  \sum_{f=u,d,s} \frac{|q_f| B \, T}{(2 \pi)^2} \,  \sum_{\nu = 0}^{\infty}  (2-\delta _{0 \nu })    \int_0^{\infty} dp_z \nonumber \\ && \hspace*{5mm} 
\left\{ \ln \left[ 1+3\left(\phi+\phi^* e^{-\frac{E_{B, f} -\mu _f}{T}}\right)\; e^{-\frac{E_{B, f} -\mu _f}{T}} +e^{-3 \frac{E_{B, f} -\mu _f}{T}}\right] \right. \nonumber \\ 
&& \hspace*{3.5mm} \left.+\ln \left[ 1+3\left(\phi^*+\phi e^{-\frac{E_{B, f} +\mu _f}{T}}\right)\; e^{-\frac{E_{B, f} +\mu _f}{T}}+e^{-3 \frac{E_{B, f} +\mu _f}{T}}\right] \right\}. \label{PloykovPLSM}
\end{eqnarray}

\begin{figure}[htb]
\centering{
\includegraphics[width=4.2cm,angle=-90]{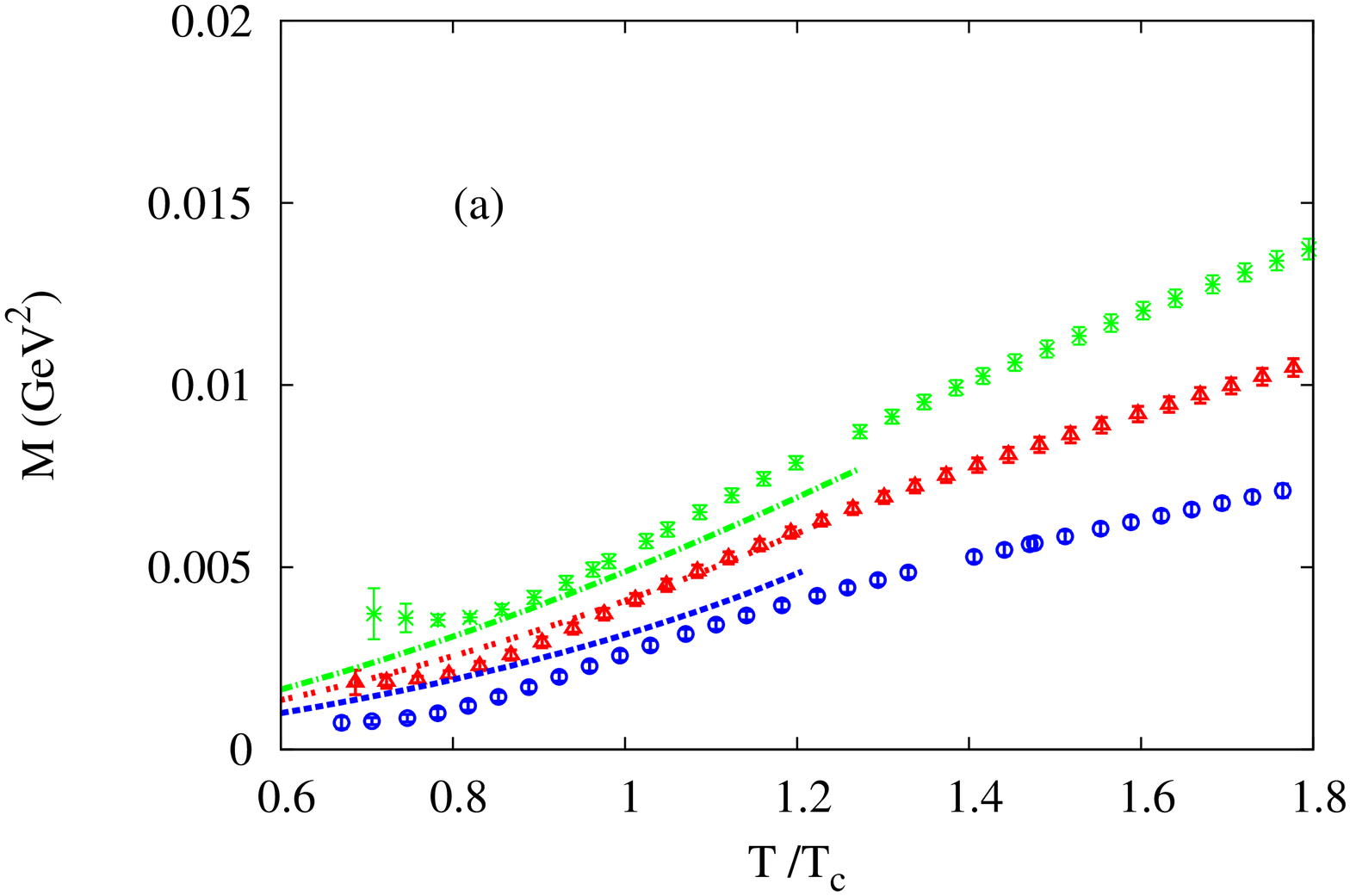}
\includegraphics[width=4.2cm,angle=-90]{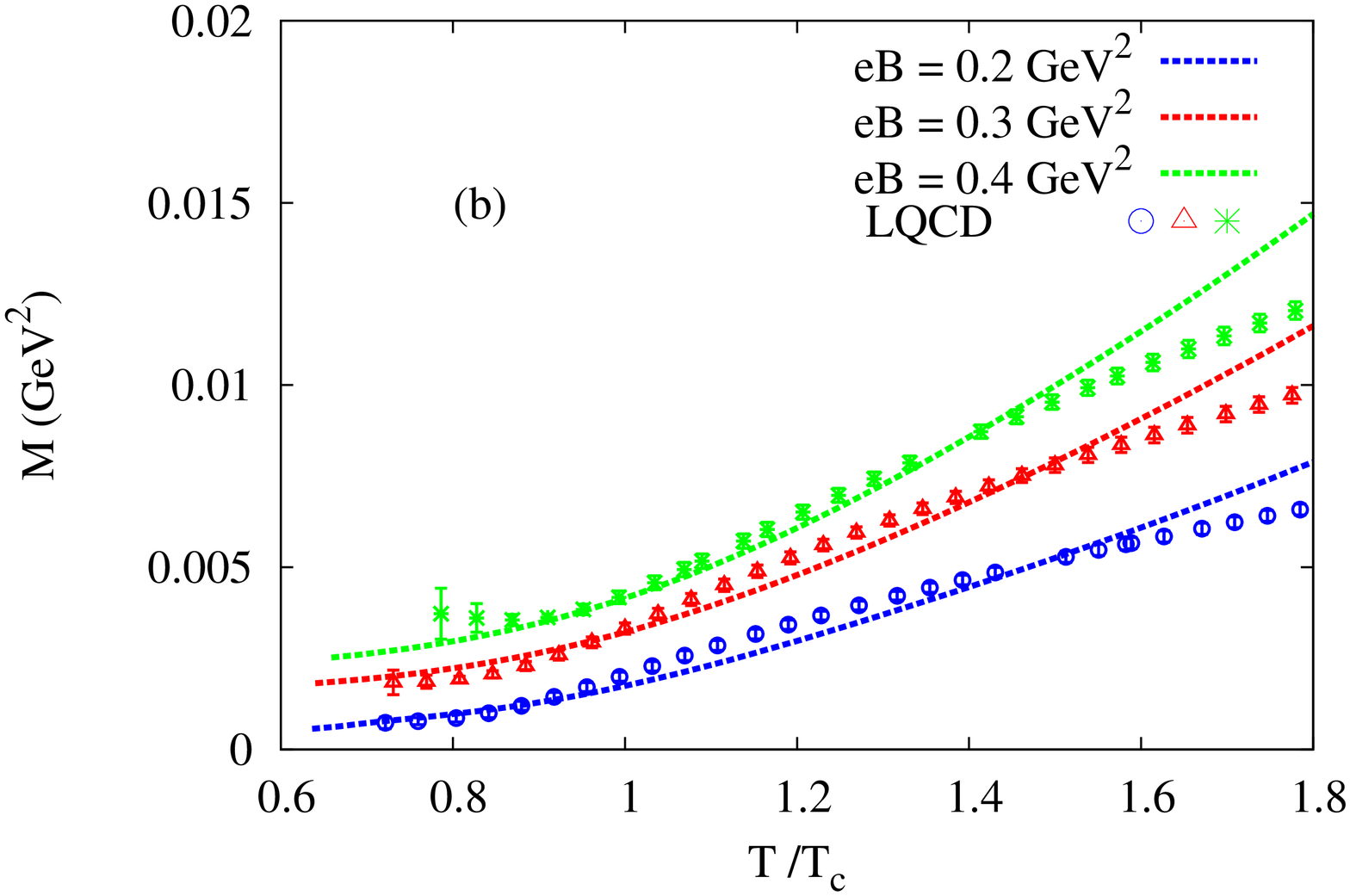}\\
\includegraphics[width=4.cm,angle=-90]{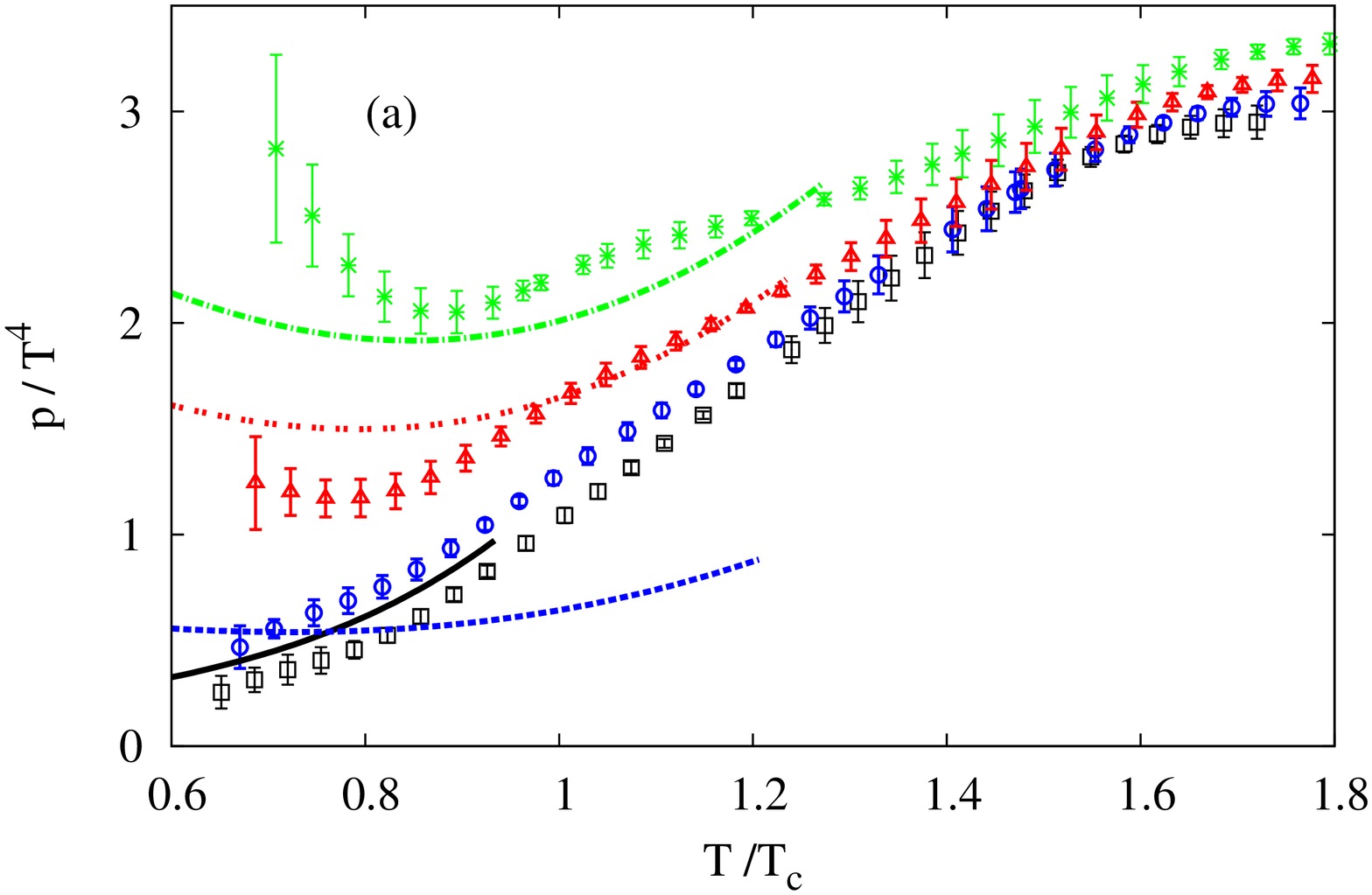}
\includegraphics[width=4.cm,angle=-90]{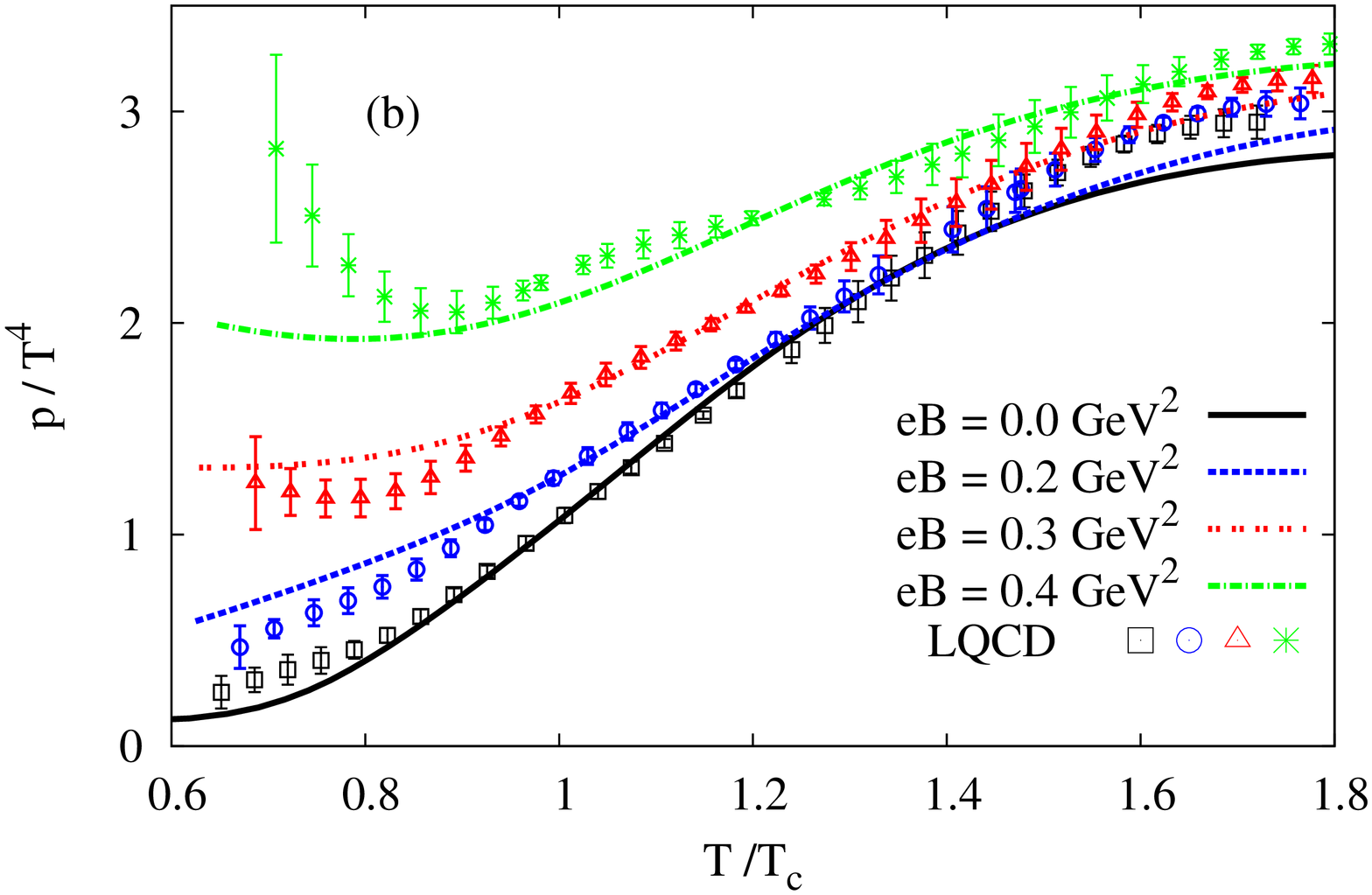}
\caption{\footnotesize The temperature dependence of the magnetization and pressure is calculated from HRG and from PLSM at different magnetic fields, $eB=0.0-0.4 $ GeV$^{2}$ (curves) and compared with recent lattice QCD calculations  (symbols) \cite{lattice3}. 
\label{fig:magn}}
}
\end{figure}

The magnetization gives one of the most fundamental properties of the QCD matter $\mathcal{M}=(T/V) \partial \ln{\mathcal{Z}}/\partial \, eB$. where $e \neq 0$ is the elementary electric charge. Positive magnetization refers to para-, while negative $\mathcal{M}$ to diamagnetism. Top panel of Fig. \ref{fig:magn} presents the magnetization calculated form HRG model and PLSM  at different values of magnetic fields as function of temperature (curves). The results are confronted  to recent lattice QCD simulations (symbols) \cite{lattice:2014}. The agreement is fairly good. We conclude that the QCD matter possesses paramagnetic property for a wide range of temperature. The bottom panel illustrates the thermodynamic pressure. Again the agreement between HRG and PLSM calculations and lattice QCD simulations is good. 

\end{document}